# A Mathematical Negotiation Mechanism for Distributed Procurement Problems and a Hybrid Algorithm for its Solution

**Zohreh Kaheh[1], Reza Baradaran Kazemzadeh[1*], Ellips Masehian[2], Ali Husseinzadeh Kashan[1]**

[1] Faculty of Industrial and Systems Engineering, Tarbiat Modares University, Tehran, Iran.
[2] Industrial and Manufacturing Engineering Dept., California State Polytechnic University, Pomona, CA, USA.

*Abstract*—In this paper, a mathematical negotiation mechanism is designed to minimize the negotiators' costs in a distributed procurement problem at two echelons of an automotive supply chain. The buyer's costs are procurement cost and shortage penalty in a one-period contract. On the other hand, the suppliers intend to solve a multi-period, multi-product production planning to minimize their costs. Such a mechanism provides an alignment among suppliers' production planning and order allocation, also supports the partnership with the valued suppliers by taking suppliers' capacities into account. Such a circumstance has been modeled via bi-level programming, in which the buyer acts as a leader, and the suppliers individually appear as followers in the lower level. To solve this nonlinear bi-level programming model, a hybrid algorithm by combining the particle swarm optimization (PSO) algorithm with a heuristic algorithm based on A* search is proposed. The heuristic A* algorithm is embedded to solve the mixed-integer nonlinear programming (MINLP) sub-problems for each supplier according to the received variable values determined by PSO system particles (buyer's request for quotations (RFQs)). The computational analyses have shown that the proposed hybrid algorithm called PSO-A* outperforms PSO-SA and PSO-Greedy algorithms.

## 1. Introduction

Negotiation-based procurement mechanism design has recently attracted much attention in academic studies. Typically, real-world procurement problems emerge as the negotiation mechanism in decentralized circumstances, in which decision-makers take action in a hierarchical structure. The multi-level programming methods are developed to solve the decentralized problems with multiple decision-makers in a hierarchical structure. The bi-level programming problem (BLPP) is a special case of multi-level programming problems with two levels of decision-makers. In the BLPPs, each decision-maker tries to optimize its own objective function without considering the objectives of other decision-makers, yet the decision of each party affects the objective values of the other parties (Kuo and Huang, 2009 & Hejazi *et al.*, 2002).

    The reason for interdependence between two levels lies in the way of creating the inducible region for the upper-level decision-maker. The inducible region is determined through optimizing the variables of the lower-level decision-makers, based on feasible values of the upper-level variables. Eventually, the upper-level decision-maker finds its optimal variable values in this inducible region. It should be noted that the

---

* Corresponding author. (R. Baradaran Kazemzadeh). Tel.: (+98)-21-82883537, Fax: (+98)-21-88005040, E-mail: rkazem@modares.ac.ir, Postal address: Faculty of Industrial and Systems Engineering, Tarbiat Modares University, P. O. Box 14115-143, Tehran, Iran.

bi-level programming is an NP-Hard problem (Ben-Ayed and Blair, 1990 & Hejazi *et al.*, 2002), even if all the objective functions and constraints are convex. For more details about bi-level programming, the reader is referred to (Cheng *et al.*, 2019) and (Chalmardi and Camacho-Vallejo, 2019).

In the following, the main areas in the literature, including the bi-level programming for decentralized procurement problems and available solution algorithms will be briefly reviewed. The Bi-level Programming problem is a special case of multilevel programming (MLP), which is categorized as a non-convex programming problem that is NP-hard (Ben-Ayed and Blair, 1990). Several methods have been presented to solve BLPP, like methods based on Kuhn–Tucker conditions (Roghanian *et al.*, 2008), fuzzy approach (Kaheh *et al.*, 2019a, 2019b, Kaheh *et al.*, 2020), Metaheuristic algorithms like the Genetic Algorithm (GA) (Hejazi *et al.*, 2002), PSO (Jung and Do Chung, 2020; Soares *et al.*, 2019), and hybrid Metaheuristics (Kuo and Han, 2011).

Recently, the PSO algorithm has been used for different bi-level programming problems (Jiang *et al.*, 2013). In many papers, it has been shown that the PSO algorithm, especially in combination with the search method, is more effective and efficient than the evolutionary algorithms such as GA. For example, Kuo and Huang (2009) provided a PSO-based method for BLPPs and used it for solving the four simple problems in the supply chain. They illustrated that the PSO algorithm outperforms GA for most of the problems. Kuo and Han (2011) develop a method based on a hybrid of GA and PSO for BLPPs. Wan *et al.* (2013) presented a hybrid algorithm by combining the PSO with a chaos searching technique for solving the nonlinear bilevel programming problems, which was more effective than an evolutionary algorithm. Soares *et al.*, 2019 modeled the interaction between a retailer and consumers as a bi-level programming problem. However, if the lower level problem, which deals with the optimal operation of the consumer's appliances, is difficult to solve, it may not be possible to obtain its optimal solution. They proposed a hybrid PSO for the mixed integer programming (MIP) model to estimate good quality bounds for the upper level (UL) objective function. Jung and Do Chung (2020) developed a bi-level PSO model as a robust optimization approach to solve the aggregate production planning problem for maintaining the effectiveness and responsiveness of manufacturing and supply chain systems.

The most relevant study to this paper is Ocampo *et al.* (2021); they developed a game-theoretic model to analyze a single manufacturer-multiple supplier, multi-period, make-to-order supply chain. They assumed that the supply chain faces a price and lead-time-sensitive demand, which is common in a make-to-order production. The vertical interaction within the supply chain is played as a Stackelberg game, where the manufacturer is considered the leader and the suppliers as the followers. A brief literature review including this paper's contributions in the last row is presented in Table 1.

This paper concentrates on solving a real-world procurement problem, where the partners intend to maintain their valuable partnerships according to their objectives in a distributed situation. This study aims at developing bi-level programming to deal with a negotiation-based procurement problem, according to realistic assumptions. Therefore, Procurement planning and order allocation to the suppliers are modeled via bi-level programming, in which the buyer is considered as a leader and makes optimal decisions according to suppliers' proposals, and the suppliers (as followers) act following the leader's decisions. Such a mechanism provides an alignment among suppliers' production planning and order allocation, to avoid the instantaneous orders, suppliers' inability to supply the orders, and imposing the high inventory cost. Besides, it supports the partnership with valued suppliers through suitable order allocation by considering suppliers' capacities.

Table 1. Summary of the reviewed literature.

| Reference | Mathematical Modeling | | | | | | | | | | | | | | | | | | Solution Approach | Real-world Case Study |
|---|---|---|---|---|---|---|---|---|---|---|---|---|---|---|---|---|---|---|---|---|
| | Main Problem | Objective Function | Negotiation Protocol | Partners | | | | Conditions | | | | Considered Costs | | | | | Type | Decision Levels | | |
| | | | | Supplier | Manufacturer | Retailer / Distributor / Transporter | Customer / Buyer | # of Products (items) | # of Negotiations | # of Planning Periods | # of vehicles | Production | Setup | Delay | Inventory / Holding | Delivery / Transport | | | | |
| Renna and Argoneto, 2010 | Production planning + automated negotiation | Max. supplier profit | – | M | – | – | M | M | M | S | – | ✓ | – | ✓ | – | – | MILP | 1 | Discrete-event simulation; agent-based approach | e-marketplace |
| Kim and Cho, 2010 | Allocation | Min. total costs of suppliers and manufacturers | – | M | M | – | – | M | S | – | – | ✓ | – | ✓ | – | – | ILP | 1 | Exact/ CPLEX | Theoretical example |
| Cheng, 2011 | Order allocation | Min. manufacturer's cost and delay; Max. suppliers' profit; Min. difference between lead-time and actual delivery time | Reverse auction | M | – | – | S | S | S | M | – | ✓ | – | ✓ | – | – | LP | 2 | Fuzzy Max-Min | Theoretical example |
| Lang and Fink, 2012 | Automated negotiation | Max. total profit | Combinatorial auctions | – | – | – | M | M | S | – | – | – | – | – | – | – | ILP | 1 | GRASP & Greedy algorithms | Theoretical example |
| Ma et al., 2013 | Pricing + Lot sizing | Max. profits for manufacturer and retailer | – | – | S | S | – | S | S | S | – | ✓ | – | ✓ | – | – | ILP | 1 | Hybrid PSO and Differential Evolution | Theoretical example |
| Baradaran Kazemzadeh et al., 2014 | Order Replenishment + Production planning | Min. suppliers' costs | Not mentioned | M | – | – | – | M | – | M | M | ✓ | ✓ | ✓ | ✓ | ✓ | MINLP | 1 | A* search | Automotive parts supplier |
| Jia et al., 2016 | decentralized planning approach: Production planning + Transportation | Min. deviations from the pickup plan received from transport operator; Min. inventory levels; Max. profit (transportation revenue minus distance-related cost, product-related cost, planning change penalties due to late and early pickup, and the external resource cost) | Distributed | – | M | M | M | M | S | M | M | ✓ | – | ✓ | ✓ | ✓ | LP | 1 | Defining negotiation space and relaxation degrees | One manufacturer, one transport operator, and two customers |

| Reference | Mathematical Modeling | | | | | | | | | | | | | | | | | | Solution Approach | Real-world Case Study |
|---|---|---|---|---|---|---|---|---|---|---|---|---|---|---|---|---|---|---|---|---|
| | Main Problem | Objective Function | Negotiation Protocol | Partners | | | | Conditions | | | | Considered Costs | | | | | Type | Decision Levels | | |
| | | | | Supplier | Manufacturer | Retailer / Distributor / Transporter | Customer / Buyer | # of Products (items) | # of Negotiations | # of Planning Periods | # of vehicles | Production | Setup | Delay | Inventory / Holding | Delivery / Transport | | | | |
| Porch et al., 2017 | Optimal control model characterizing the decision to invest in supplier development | Max. profit for supplier & manufacturer | – | S/M | S | – | – | S | S | S | – | ✓ | – | – | – | – | – | 1 | a negotiation-based algorithm using differentiation | Theoretical example |
| Shokr and Torabi, 2017 | Procuring excess required relief items for humanitarian organizations at post-disaster | Max. profit; Min. procurement costs; Min. maximum delivery times | Enhanced reverse auction | M | – | – | – | M | S | M | M | – | – | ✓ | – | ✓ | Probabilistic MINLP | 1 | An iterative possibilistic approach | An illustrative numerical example based on the Iranian Red Crescent Society |
| Soares et al., 2019 | Dynamic tariffs in the electricity retail market | Max. retailer's profit; Min. consumer's electricity bill | – | – | – | S | M | – | – | M | – | ✓ | – | – | – | – | MILP | 2 | Hybrid PSO | Electricity retail market |
| Zhang and Wang, 2019 | Network design + Transfer-pricing | Max. total profit | – | S | S | S | – | M | S | M | M | ✓ | – | – | ✓ | ✓ | MILP | 1 | N/A | Subsidiaries in Japan, China, and the United States |
| Ocampo et al., 2021 | Make-to-order manufacturing supply chain planning | Max. system-wide profit | – | M | S | – | – | M | S | M | – | ✓ | ✓ | ✓ | ✓ | ✓ | LP | 2 | Stackelberg game | Steel industry |
| Current study | Decentralized planning approach: Distributed procurement problem+ Order allocation + Production planning | Min. procurement cost and delay penalty | Reverse auction and bidding/ distributed | M | S | – | – | M | S | M | M | ✓ | ✓ | ✓ | ✓ | ✓ | MINLP | 2 | Hybrid PSO-A* algorithm | Automotive parts supplier |

*Legend*: S = Single, M = Multiple, LP = Linear programming, ILP = Integer linear programming, MILP = Mixed-integer linear programming, MINLP = Mixed-integer nonlinear programming.

As Table 1 shows, the majority of the relevant works on Procurements Problems have proposed single-level models to solve the problem. However, the problem's decision-making process has a hierarchical nature and single-level models are not capable of satisfying/optimizing the interests of all the partners involved in the transactions. Therefore, in this regard, bi-level models have an apparent advantage over single-level models. Our proposed approach can deal with simultaneous interactions between the buyer and each supplier in a one-period contract and captures the hierarchical nature of the decision-making process. Our lower-level model is a set of mixed-integer nonlinear programming (MINLP) subproblems with a discrete space where Karush-Kuhn-Tucker conditions (KKT) cannot be applied. Using a centralized (single-level) model or the KKT optimality conditions (Carrión, 2009 and Moiseeva and Hesamzadeh, 2017) incorrectly eliminates the distributed nature of the problem and creates the wrong impression that the buyer directly influences the suppliers' production planning (the followers' decision variables) toward its own interests.

In addition, unlike other bi-level works reviewed in Table 1, our proposed model considers the most comprehensive set of costs, as well as accommodates multiple products (items), negotiations, and the number of vehicles. These assumptions, together with the proposed novel hybrid solution mechanism, provide the main contributions of this paper.

This paper proposes an innovative hybrid algorithm based on the PSO algorithm, in which an A* search is applied to solve the nonlinear mixed-integer programming subproblems for each supplier in the lower level. Applying a metaheuristic algorithm is well-founded owing to some attributes such as the inherent complexity of the bi-level programming problems, the discrete inducible region, and the abundance of integer and binary variables. Furthermore, applying a population-based algorithm is preferable because our problem deals with plenty of local solutions as agreement points on the inducible region which makes it difficult to detect the buyer's optimal solutions. Among population-based algorithms, the PSO algorithm has many advantages such as short run time and less memory requirement (Kadadevaramath et al, 2012). These properties are especially important when the PSO algorithm is combined with the A* search.

In the proposed model, the leader employs a hybrid particle swarm optimization algorithm as a decision-making strategy to make an alignment among separate transactions and to achieve a near-optimal solution. Decision-making in the lower level is done according to the received variable values frequently determined by PSO particles. Through embedding the A* search for each supplier, the suppliers in the lower level model are considered as problem-solving entities whose decision-making strategy is based on the A* search.

## 2. Mathematical Modeling of the Negotiation Mechanism

The distributed procurement problem is modeled through bi-level programming, in which the buyer is considered as the leader, and suppliers are considered as independent followers. The lower-level model is a set of subproblems, which are multi-period and multi-product production planning problems for each supplier, while the upper level is a one-period contracting, in which the buyer aims to procure a bundle of similar items. The upper-level decision-maker intends to minimize the procurement cost and shortage cost due to the suppliers' delays in delivery concerning the requested due date. The buyer determines the quantity of the allocation to each supplier and two predetermined due dates, including early and late acceptable due dates. Meanwhile, the suppliers are aware of the early due date, and their delay penalty is calculated based on it. On the other hand, the buyer shortage cost is calculated based on the late due date, which means the buyer may have encountered the shortage after this time. In the lower level, each supplier makes decisions separately about items' prices and quantities in each delivery and announces them to the upper level.

The detailed description of our proposed model is as follows: first, the upper-level decision-maker allocates orders to each supplier to satisfy its demand for each item. According to the allocated quantities to each supplier, supplier *i* compares its available inventory with the order quantity for item *j*. If the allocated quantity is below the inventory, the orders will be sent from the warehouse; otherwise, the production line must be launched to meet the remaining demand for item *j*. Working periods (including ordinary time and overtime), as well as the upper bound of production capacity for each supplier in each period (considering the processing time of each item), are definite. The production rate in ordinary time and overtime is similar, but their production costs are different. The manufactured items are kept in the warehouse, so each supplier deals with the capacitated inventory and inventory cost during the working period and in the interval between the two consecutive working periods. The delivery cost is calculated according to a fixed cost for each used truck, and a variable cost, which is a product of a determined coefficient and the loaded items. The lower-level decision-makers aim to minimize their total cost, including production cost, delivery cost, inventory cost, setup cost, and delay penalty according to the allocated items, and the due date stated by the buyer. In this way, the delivery cost implicitly is considered in bid prices. Delay penalties for suppliers only reduce the supplier credit and do not affect the price, and consequently do not lead to any kind of benefit for the buyer. It should be noted that the delay cost is a product of the number of items sent with a delay and the time delay, which is adjusted by a coefficient. Each of the suppliers, according to the total cost minimization and the lowest acceptable interest rate of profit, announces items' prices and quantities for each delivery to the buyer. The model assumptions are listed as follows:

1. Only one contract period is considered.
2. The orders are delivered instantaneously to the buyer in several steps and during the periods.
3. The suppliers' production capacities are different.
4. The production cost of each item includes setup cost, but the setup time is not considered.
5. Each period includes ordinary working time and overtime, and production cost is higher in overtime.
6. The delivery cost depends on vehicle rental and the number of loaded items.
7. There are finite numbers of available vehicles in each period.
8. The inventory is considered for finished items, but not for semi-finished items. Inventory cost depends on the number of items at the unit of time.
9. The buyer determines a pair of due dates and the suppliers are aware of the earlier due date, and calculate their delay cost based on it. However, the buyer's delay cost is calculated based on the later due date.
10. The suppliers' delay cost is not profitable for the buyer.
11. The suppliers are not aware of each other's proposals.
12. The raw material price for all suppliers is a fixed and definite parameter so it is not considered.
13. An initial safety stock for each type of finished item is considered at the beginning of the time horizon.

## 2.1. The Bi-Level Programming Model

In this section, the proposed mathematical model will be explained. The parameters and decision variables are given below. The set $S_1 = \{q_{ij}; x_{ij}; updc_{ij}^t\}$ are the variables which are controlled by the buyer in the upper level, and the set $S_{i2} = \{p_{ij}; yr_{ij}^t; yn_{ij}^t; send_{vij}^t; lodc_{ij}^t; x'_{ijt}; x''_{vijt}; I_{ij}^t\}$ denotes the variables that are controlled by each supplier at the lower level. The variables exchanged between the two levels include $\{q_{ij}; p_{ij}; send_{vij}^t\}$. The mathematical model will be presented after defining the variables and parameters.

- Indices:
    - $t$: period's index, $t = 0, \ldots, T$
    - $i$: supplier's index, $i = 0, \ldots, n$
    - $j$: item's index, $j = 0, \ldots, m$
    - $v$: vehicle's index, $v = 0, \ldots, V$

- Decision variables in the upper level:
    - $q_{ij}$: The allocated quantity of item $j$ to supplier $i$ (an integer variable).
    - $x_{ij}$: The binary variable for allocating the item $j$ to supplier $i$.
    - $updc_{ij}^t$: The shortage cost for buyer due to the delay of supplier $i$ to deliver the item $j$.

- Parameters in the upper-level model:
    - $D_j$: Buyer's total demand for item $j$.
    - $(LT_{lower}; LT_{upper})$: The due date range suggested by the buyer to all suppliers.
    - $(Q_{ij}^{min}; Q_{ij}^{max})$: The upper bound and lower bound for allocating the item $j$ to supplier $i$.
    - $\lambda$: Delay adjustment factor for the buyer
    - $a_{ij}$: ordering cost to supplier $i$ for item $j$.

- Decision variables in the lower-level model:
    - $p_{ij}$: The $j$-th item's price offered by supplier $i$.
    - $x'_{ijt}$: Binary variable for producing the item $j$ by supplier $i$ in period $t$.
    - $x''_{vijt}$: Binary variable for delivering the item $j$ by vehicle $v$ from supplier $i$ in period $t$.
    - $yr_{ij}^t$: The production volume of the item $j$ by supplier $i$ in ordinary time in period $t$.
    - $yn_{ij}^t$: The production volume of the item $j$ by supplier $i$ in overtime in period $t$.
    - $send_{vij}^t$: The integer variable for the volume of item $j$ in each delivery through vehicle $v$ by supplier $i$ in period $t$.
    - $I_{ij}^t$: Inventory of item $j$ at the end of the $t$-th period for supplier $i$.
    - $lodc_{ij}^t$: The penalty for supplier $i$ due to the delay of item $j$ in period $t$.
    - $TC_{ij}$: Total cost for supplying the item $j$ by supplier $i$.

- Parameters in the lower-level decision model:
    - $cor_{ij}$: The production cost during ordinary time
    - $cov_{ij}$: The production cost during overtime
    - $orc_{ij}$: Production capacity in ordinary time
    - $ovc_{ij}$: Production capacity in overtime
    - $PT_{ij}$: Processing time
    - $H_{ij}$: Inventory cost for each unit of item $j$ by supplier $i$ during each period
    - $H'_{ij}$: Inventory cost for each unit of item $j$ by supplier $i$ in the time interval between two periods
    - $sc_{ij}$: The setup cost for producing the item $j$ by supplier $i$
    - $ss_{ij}$: The safety stock of the item $j$ for supplier $i$

$VCap_{ij}$: the capacity of the $i$-th supplier's vehicle for delivering the item $j$

$InCap_{ij}$: the capacity of the $i$-th supplier's warehouse for delivering the item $j$

$\gamma$: Delay adjustment factor for the suppliers

$g_i$: Acceptable profit rate for supplier $i$

- The Buyer's decision-making model (upper-level decision-maker) is:

$$\min_{S1} Z = w_1 \cdot \sum_{i=1}^{n}\sum_{j=1}^{m}(p_{ij}q_{ij} + x_{ij}a_{ij}) + w_2 \cdot \sum_{i=1}^{n}\sum_{j=1}^{m}\sum_{t=1}^{T} x_{ij} \cdot updc_{ij}^{t} \quad (1)$$

S.T.:

$$\sum_{i=1}^{n} q_{ij} = D_j \qquad \forall j \quad (2)$$

$$x_{ij}Q_{ij}^{\min} \leq q_{ij} \leq x_{ij}Q_{ij}^{\max} \qquad \forall i,j \quad (3)$$

$$\lambda(t - LT_{upper}) \cdot \sum_{v=1}^{V} send_{vij}^{t} \leq updc_{ij}^{t} \qquad \forall i,j,t \quad (4)$$

$$x_{ij} = \{0,1\} \qquad \forall i,j \quad (5)$$

$$q_{ij} \geq 0, \text{int}, \ updc_{ij}^{t} \geq 0 \qquad \forall i,j,t \quad (6)$$

- The Suppliers' decision-making model (a set of nonlinear mixed-integer programming subproblems in lower level) is:

$$\underset{S_i 2}{Min} = \{Z_1, Z_2, ..., Z_i \mid Z_i = \sum_{j=1}^{m} TC_{ij}; i = 1,...,n\} \quad (7)$$

S.T.:

$$TC_{ij} = \sum_{t=1}^{T}\left[\begin{array}{l} cor_{ij} \cdot yr_{ij}^{t} + cov_{ij} \cdot yn_{ij}^{t} + H_{ij} \cdot PT_{ij} \cdot \frac{1}{2}\left(\sum_{v=1}^{V}(send_{vij}^{t})^{2} + (I_{ij}^{t})^{2} - (I_{ij}^{t-1})^{2}\right) + H'_{ij}I_{ij}^{t} + \\ \sum_{v=1}^{V}\left(\alpha(x''_{vijt}) + \beta(send_{vij}^{t})\right) + lodc_{ij}^{t} + sc_{ij}x'_{ijt} \end{array}\right] \quad \forall i,j \quad (8)$$

$$yr_{ij}^{t} \leq \frac{orc_i}{PT_{ij}} \qquad \forall i,j,t \quad (9)$$

$$yn_{ij}^{t} \leq \frac{ovc_{it}}{PT_{ij}} \qquad \forall i,j,t \quad (10)$$

$$yr_{ij}^{t} + yn_{ij}^{t} \leq M \cdot x'_{ijt} \qquad \forall i,j,t \quad (11)$$

$$\sum_{t=1}^{T} yr_{ij}^{t} + yn_{ij}^{t} = q_{ij} \qquad \forall i,j \quad (12)$$

$$I_{ij}^{t} = yr_{ij}^{t} + yn_{ij}^{t} + I_{ij}^{t-1} - \sum_{v=1}^{V} send_{vij}^{t} \qquad \forall i,j,t \quad (13)$$

$$I_{ij}^{t} \leq InCap_{ij} \qquad \forall i,j,t \quad (14)$$

$$\sum_{t=1}^{T}\sum_{v=1}^{V} send_{vij}^{t} = q_{ij} \qquad \forall\, i,j \qquad (15)$$

$$\gamma(t - LT_{lower})\sum_{v=1}^{V} send_{vij}^{t} \leq lodc_{ij}^{t} \qquad \forall\, i,j,t \qquad (16)$$

$$send_{vij}^{t} \leq x''_{vijt} \cdot VCap_{ij} \qquad \forall\, i,j,t,v \qquad (17)$$

$$send_{vij}^{t} \leq InCap_{ij} \qquad \forall\, i,j,t,v \qquad (18)$$

$$\sum_{v=1}^{V} x''_{vijt} \leq V \qquad \forall\, i,j,t \qquad (19)$$

$$p_{ij} = \frac{(1+g_i)}{q_{ij}}\left(TC_{ij} - \sum_{t=1}^{T} lodc_{ij}^{t}\right) \qquad \forall\, i,j \qquad (20)$$

$$yr_{ij}^{t}, yn_{ij}^{t}, send_{vij}^{t}, I_{ij}^{t} \geq 0, \text{int} \qquad \forall\, i,j,t \qquad (21)$$

$$lodc_{ij}^{t} \geq 0 \qquad \forall\, i,j,t \qquad (22)$$

$$x'_{ijt}, x''_{vijt} = \{0,1\} \qquad \forall\, i,j,t,v \qquad (23)$$

$$i \in \{1,...,n\}, \quad j \in \{1,...,m\}, \quad t \in \{1,...,T\}, \quad v \in \{1,...,V\} \qquad (24)$$

The buyer's objective function at the upper level is presented in (1), which aims to minimize the procurement cost and the delay penalty cost. The corresponding weight for each objective is determined through the interviews with the experts in a large supplying automobile parts corporation. As the delay cost is more important than the procurement cost, its corresponding weight has to be higher. Eq. (2) shows the total allocated quantities for each item should not be above the total demand of the buyer for the item. Eq. (3) states that the order of each item allocated to a particular supplier must be lower than or equal to a specific maximum order quantity and greater than or equal to a certain minimum order quantity. The lower bound and upper bound of allocation is determined based on the business partnership history, a guess about the level of suppliers' satisfaction with the allocated quantity, qualifications' grade for each item, and suppliers' technological capability. Eq. (4) explains the delay cost is a product of the number of items sent with a delay and the time delay in comparison with the latest acceptable buyer due date. Eq. (5) expresses that $x_{ij}$ is a binary variable to accept or reject the supplier's proposal. Eq. (6) shows that $q_{ij}$ is a non-negative integer variable.

In the lower level, each supplier individually computes appropriate bid prices for allocated items while minimizing its total costs. In doing so, their objective is to minimize their total costs separately (7). Eq. (8) computes the total cost according to the ordinary and overtime production cost, setup cost, delay cost, inventory cost, and delivery cost. Eq. (9, 10) shows the ordinary and overtime production capacities. Eq. (11) shows if production occurs, the fixed setup cost will be considered. Eq. (12) shows the supplier must satisfy the allocated quantity and leave a safety stock for the next time horizon. Eq. (13) shows the inventory equilibrium equation. Eq. (14) shows the inventory capacity for each finished item type. Eq. (15) shows the delivered quantities of each item are equal to the buyer's order. Eq. (16) explains the delay penalty calculation for each supplier and each item. Eq. (17) shows the constraint related to using a vehicle or not, according to the vehicle capacity and delivered quantity for each item. Eq. (18) considers the inventory capacity before delivery. It means the volume of item *j* in each delivery through vehicle *v* by supplier *i* in period *t* should be less than the amount of $InCap_{ij}$, which is the capacity of *i*-th supplier's warehouse for delivering the item *j* Eq. (19) shows the constraint of available vehicles in each period. Eq. (20) shows the acceptable unit price is a product of the least acceptable profit and total cost (without considering the delay cost) for each item. Eq. (21) shows the non-negative integer variables. Eq. (22) shows the non-negative variables, and Eq. (23) shows the binary variables.

It should be noted that the number of periods in a finite time horizon is estimated according to:

$$Max\left\{\frac{order\ quantity \times prossesing\ time}{ordinary\ capacity\ of\ time}; \frac{order\ quantity}{number\ of\ vehicles \times \min\{vehicle\ capacity; inventory\ capacity\}}\right\} \quad (25)$$

In the next section, two types of negotiation protocols in bi-level programming will be compared.

## 2.2. A comparison between two types of negotiation protocols in bi-level programming

Despite the availability of mathematical models for the negotiation mechanisms in the literature (e.g. Jung *et al.*, (2008) and Cheng (2011)), there has not been a strict definition of the negotiation structure in mathematical models. Two important components in each negotiation, including protocol and strategy; negotiation protocol provides clear rules to conduct the interactions of the negotiating parties that it has to be apparent to all the parties. On the other hand, Negotiation strategy is the way in which each party decides to attain the best outcome of the negotiation (Fang and Wong, 2010). There are three categories of negotiation protocols – bidding, auction, and bargaining.

To clarify the type of negotiation protocol in our procurement problem, two types of negotiation protocols that can be captured through bi-level programming will be compared. To this end, a relation between the negotiation concepts and the bi-level programming problem can be mapped.

Generally, in negotiations, the auction protocol is ordinarily used when these conditions are satisfied: (1) exactly one issue (price) to be considered, (2) does not need two-way communications between parties, and a party precisely decides based on received proposals, and (3) It is not necessary to exercise different negotiation strategies with different partners. On the other hand, the bargaining protocol is usually applied when these conditions are satisfied: (1) both sides can offer (two-way communication), (2) multiple issues can be included. Therefore, our proposed negotiation mechanism based on bi-level programming has the third property of auctioning and the first and second properties of bargaining.

Unlike our bi-level negotiation mechanism, in auction-based (reverse auction) negotiations, suppliers specify the acceptable quantities and prices and reform their proposals to maximize the winning probability and minimize their total cost. Additionally, the buyer as an auctioneer just accepts or rejects the proposals. To illustrate this matter, Sandholm *et al.* (2002) defined a reverse auction in which the auctioneer specifies the whole demand for each item; and sellers submit a set of requests, which each of them includes the number of the items and price for the bundle of requested items in a combinatorial auction. Therefore, in auction-based negotiation, the buyer's total demand is apparent for all suppliers, and the time horizon needs to be fixed by the buyer, also the delay is not allowed.

Nevertheless, in our problem, the buyer determines the amounts of order allocations and frequently reallocates to achieve an approximate optimal solution for the bi-level programming model. According to this explanation, although the buyer will not offer the prices, the applied protocol is a combination of bargaining and reverse auction. Finally, in this simulated negotiation, the buyer's strategy is a hybrid PSO-A* algorithm, which deals with a bi-level programming model. Moreover, the suppliers' strategy is the A* search, which searches the desired state for each supplier's production planning.

## 3. The proposed Solution Procedure in a Distributed System

In this paper, a hybrid algorithm is implemented to resolve the bi-level procurement problem. In regular bi-level programming, in which the lower-level model is continuous, it is simple to apply the KKT optimality conditions and convert to a centralized single-level model. Furthermore, in this condition, if there are several decision-makers at the lower level, it is possible to overcome this decentralization by applying the KKT optimality conditions. However, this action is not applicable in our proposed model because the lower model is a set of nonlinear mixed-integer programming problems. Furthermore, this paper strongly emphasizes the system distribution, so the proposed algorithm is supposed to capture the decentralized nature of the problem.

As previously mentioned, the upper level (leader) controls the transaction. In other words, the leader must be able to change its position toward its objective, according to the followers' decisions, and finds its near-optimal solution in the inducible region. It should be stated that the followers' proposals are set according to the leader's variable values (particles in the PSO population). Therefore, through implementing Particle Swarm Optimization (PSO), it is rational to say the particle swarm intelligence makes a simultaneous negotiation with individual suppliers. The reason for this assertion is that each particle in the PSO algorithm is considered as the buyer's *Request For Quotations* (RFQs), which are sent to the suppliers. Moreover, according to the constructive communication between particles in PSO as a swarm intelligence algorithm, the buyer's RFQs in each iteration is a combination of earlier RFQs with some changes (update the velocity and the position of each particle) in comparison with the prior iteration.

Also, as previously mentioned, the lower-level problem is a set of MINLP sub-problems that makes the problem harder. It takes a long time to find a solution through an exact problem solver. The reason for its computational complexity is the plethora of integer and binary variables. Thus, a heuristic algorithm based on A* search is embedded in the metaheuristic algorithm. By incorporating an A* search for each supplier, the suppliers in the lower level model are considered as problem-solving agents. The problem-solving agents are applied to deal with the decentralization at the lower level. Each problem-solving agent according to the buyer's RFQs (particles in the PSO population) attempts to solve its problem through an A* search.

The details of a novel hybrid PSO-A* algorithm for solving the proposed BLPP model will be presented in the next section of the paper. The overall steps to solve our proposed bi-level problem is expressed as follows:

*Step 1*. (*Upper-level problem*): Generate a set of feasible solutions for upper-level decision variables (population particles in PSO algorithm) as the leader's RFQs (allocated quantity to each supplier).

*Step 2*. (*Lower-level problem*): Optimize the followers' actions for each initial value of the upper-level decision variable, and return the optimal or near-optimal reactions to the leader's model. (The subproblems for each supplier are solved through an A* search, and the value of its variables is returned)

*Step 3*. (*Upper-level problem*): Evaluate the leader's objective value for the variable values in the upper level and corresponding values in the lower level.

*Step 4.* (*Upper-level problem*): If the termination condition has not been met, the leader's variable values will move to new positions and go to step 2 until a proper stop criterion is met and an optimal or near-optimal solution is achieved.

The details of the PSO algorithm for the bi-level problem and the A* based heuristic algorithm will be described in Sections 4.1 and 4.2, respectively.

## 3.1. Details of the Proposed PSO algorithm

In the previous section, the overall solution procedure for our proposed bi-level programming model was introduced. Now, the details of our proposed PSO algorithm are described to solve the bi-level problem. For a general review of different versions of the PSO algorithm, the reader is referred to Sedighzadeh and Masehian, 2009, Kadadevaramath *et al.*, 2012, and Soares *et al.*, 2019. The essential reasons for applying the PSO algorithm are brought as follows:

1. Among the population-based metaheuristic algorithms especially the evolutionary algorithms, the PSO algorithm needs the shortest run time and memory requirement (Kadadevaramath *et al.*, 2012). These properties are especially important for combining the PSO algorithm with a search method (A* search) in our proposed mechanism.
2. Each particle in the PSO algorithm is considered as a buyer's RFQ which is sent to suppliers. On the other hand, it should be mentioned that the PSO algorithm, unlike the evolutionary algorithms which are based on the survival of the fittest principle, is based on the constructive cooperation among the particles, so the next buyer's RFQs is based on its best experience and the global best.

The steps of our proposed linearly decreasing inertia weight PSO algorithm are described as follows:

*Step 1 – Initialization*: The solution representation is a matrix in which each row corresponds to a supplier, and each column corresponds to an item, and also its elements represent the allocated quantities. To satisfy the demand of each item, a supplier is randomly selected and a number between ($Q_{ij}^{min}$, Min ($Q_{ij}^{max}$, remaining demand)) is assigned to its element, this action is repeated frequently to fill all elements of the initial positions' matrix (Fig. 1). Furthermore, the parameters of the algorithm, including population size (the number of particles), inertial weight, and two learning factors are also set up. The initial velocity is set to zero, and velocity intervals are determined based on a coefficient of the quantity interval length.

$$\begin{bmatrix} q_{11} & \cdots & q_{1m} \\ \vdots & q_{ij} & \vdots \\ q_{n1} & \cdots & q_{nm} \end{bmatrix}$$

**Figure 1.** The representation of PSO solution.

*Step 2. Initial personal best (P-best) and global best (G-best):* The set of all matrices in the first step (initial population particles) are local solutions and the best among them is considered as the global solution.

*Step 3. Generate the new population:* to generate the new population, the velocity and position of each particle are updated as follows:

*Step 3-1. Velocity update:* velocity of each particle update according to:

$$v_{ij}^t = wv_{ij}^{t-1} + c_1 r_1 (p_{ij}^{t-1} - x_{ij}^{t-1}) + c_2 r_2 (G_{ij}^{t-1} - x_{ij}^{t-1}) \tag{26}$$

If the new velocity is out of the default interval, its value is modified to the upper or lower bound of velocity according to:

$$v_{ij}^t = \begin{cases} v_{max} & v_{ij}^t > v_{max} \\ v_{min} & v_{ij}^t < v_{min} \end{cases} \tag{27}$$

If an initial position is zero, the zero value is replaced with $Q_{ij}^{min} - \varepsilon$ to update the position; this action causes slight changes of velocities during the final stages. It should be stated that the inertia weight is updated with a decreasing trend according to:

$$w = w_{max} - \frac{w_{max} - w_{min}}{iter_{max}} \times iter \tag{28}$$

*Step 3-2. Position update:* position of each particle update according to:

$$x_{ij}^t = x_{ij}^{t-1} + v_{ij}^t \tag{29}$$

Because of the decimal velocity vector, the new positions will be decimal. So the floor of positions is checked in demand satisfaction constraint and sent to lower-level decision-makers.

*Step 4. Demand satisfaction constraint and the new population modification*: The floor of newly generated solutions (integer solutions) may not be equal to total demand for each item, and the constraint $\sum_{i=1}^{n} q_{ij} = D_j$ be violated. So the summation of allocation solutions (new positions) is compared with the total demand for each item. If those are equal, the generated solutions do not need to be modified. If it is smaller (or larger), a supplier is randomly selected, and pluses (or subtracts) one unit to its value by taking predefined quantity interval into consideration. If the equalization is not satisfied, this action is frequently repeated to achieve equalization. It must be noted that adding a unit to a position value when its value is zero, means it is replaced with the lower bound of quantity allocated, and vice versa according to:

$$\begin{cases} x_{ij}^t + 1 = Q_{ij}^{min} & x_{ij}^t = 0 \\ x_{ij}^t - 1 = 0 & x_{ij}^t = Q_{ij}^{min} \end{cases} \tag{30}$$

After going out of the modifying loop, to prepare for the next generation the zero positions are replaced with $Q_{ij}^{min} - \varepsilon$, and the decimal part of the solution is added to the modified solution.

*Step 5. Transactions between the two levels*: The generated matrix is sent to the lower-level decision-makers, equipped with A* search, to obtain their optimized variables, then the upper-level objective function is calculated. It should be stated that the new positions are decimal because of the decimal velocity vector. So the floors of new positions (quantity values) are sent to lower-level decision-makers. By the way, it is assumed that if new positions are smaller than the lower limit, replaced with zero values before sending to the lower level.

*Step 6. P-best and G-best update:* the function value of each particle is compared with the old personal best and global best in the last iteration according to:

$$\begin{cases} P_{ij}^t = x_{ij}^t & f(x_{ij}^t) < P_{ij}^{t-1} \\ G_{ij}^t = x_{ij}^t & f(x_{ij}^t) < G_{ij}^{t-1} \end{cases} \tag{31}$$

*Step 7. Termination:* Terminate if the maximum number of iterations is reached, otherwise repeat steps 4 to 7. In the end, return the global best and its function value.

## 3.2. The A*-based Heuristic Algorithm

There are many problem-specific heuristics for mixed-integer programming problems (Akartunali and Miller, 2009 & Danna *et al.*, 2005), but barely an efficient and effective heuristic can be found. To solve the lower-level MINLP sub-problems a heuristic algorithm based on A* search will be proposed in this paper. A* search is optimally efficient for any given heuristic function, and no other optimal algorithm is guaranteed to expand fewer nodes than A* (Russell and Norvig, 1995). The interested reader is referred to (Hansen and Zhou, 2007) for more information.

Through embedding the A* search for each supplier, the suppliers in the lower level model are considered as problem-solving agents. The problem-solving agents are applied to solve the mixed-integer nonlinear programming subproblems for each supplier in the lower level. Problem-solving agents as a kind of goal-based agent have been introduced by Russell and Norvig (1995): "Problem-solving agents decide what to do by finding sequences of actions that lead to desirable states". Before describing the algorithm steps, some definitions have to be introduced. An A* search tree for a numerical example has been illustrated in Fig. 2 and the members of the Open-list and Closed-list in each period have been indicated in Table 2 (Baradaran Kazemzadeh et.al., 2014).

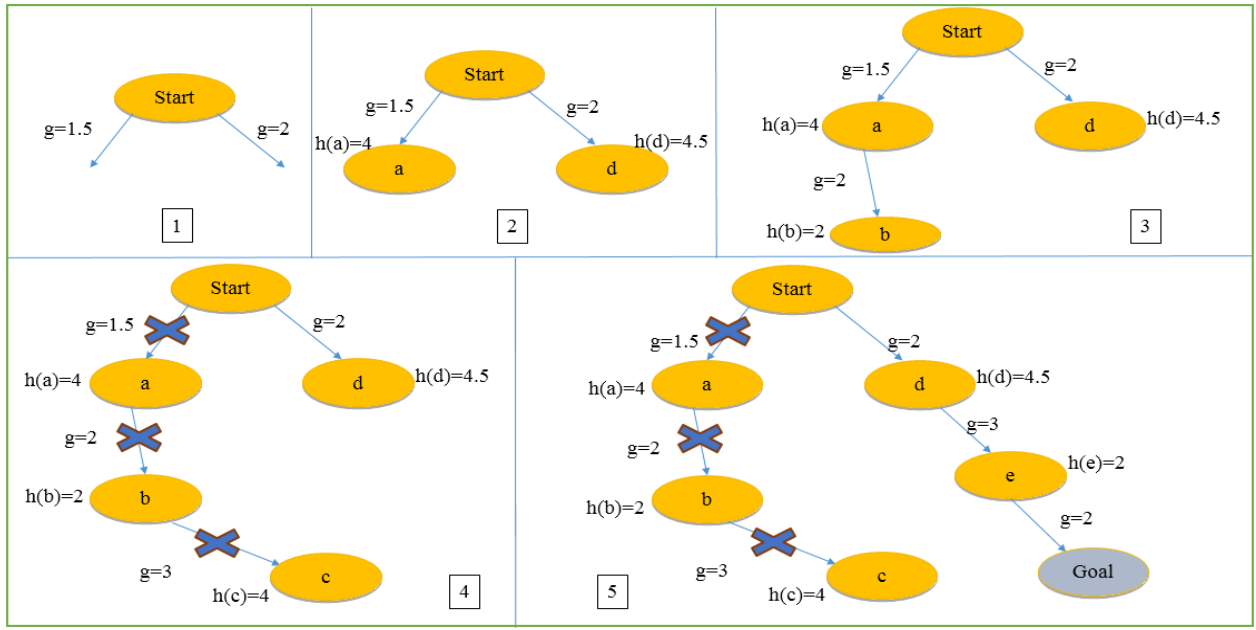

**Figure 2.** The A* search tree for a numerical example.

**Table 2.** The members of the *Open-list* and *Closed list* for the example in Figure 2.

| Period | Open-list | Closed-list |
|---|---|---|
| 1 | {start} | {} |
| 2 | {a, d} | {start} |
| 3 | {b, d} | {start, a} |
| 4 | {d, c} | {start, a, b} |
| 5 | {c, e} | {start, a, b, d} |
| 6 | {e} | {start, a, b, d, c} |
| 7 | {} | {e} |

The *initial state* in our problem is a state in which a supplier is, for the first time. Initial state attributes include the initially available inventory and the total allocated order quantity. *Operators* are any action, including ordinary time production, overtime production, and delivery volume, which change the system state and generate another state. *State-space* is the set of all states reachable from the initial state by any sequence of actions, including available inventory in the current period and remaining of allocated demand. The *Path* is any sequence of actions leading from one state to another state, such as the decision-making about production, storage, or delivery. The *goal state* is a state, in which the remaining allocated order is zero and the buyer's demand is satisfied. The *goal test* is done at the end of each period in a finite time horizon, and *path cost* is according to the lower-level objective function. Together, this initial state, operator set, goal test, and path cost function define each supplier's problem. It should be stated that the node selecting in A* search is based on combining two evaluation functions, including the $g(n)$ cost and $h(n)$ cost according to Eq. 32 (Hart *et al.*, 1968). The $g(n)$ cost gives the path cost from the start node to the current node, and the $h(n)$ cost is an estimated cost of the cheapest path from the current node to the goal.

$$f(n) = g(n) + h(n) \tag{32}$$

The A* search usually runs out of space long before it runs out of time (Russell and Norvig, 1995), so a limitation for the number of open-list members should be considered to reduce the memory requirement and overcome this difficulty. The Heuristic algorithm steps for each supplier and the calculated order quantity of each item are as follows. Since A* search is used in Step 5, we numbered them as 5.1, 5.2, and so on.

*Step 5.1.* At first, the order quantity should be compared with the initial inventory. If the inventory is enough, the order is sent from the warehouse, and its cost is based on total cost except for overtime production cost and setup cost. Then the algorithm is finished, otherwise, go to step 5.2.

*Step 5.2.* The start node is considered as the state in which the supplier faces the remaining demand, and tries to set up the production line. This node is added to the *open-list*.

*Step 5.3.* The selected node in the open-list is considered as a parent, and branched as follows:

*Step 5.3.1.* Ordinary production could be the integer numbers between [0, min (remaining demand, ordinary production capacity)].

*Step 5.3.2.* If the ordinary capacity is fully utilized, the overtime capacity will be used. The overtime production could be the integer numbers between [0, min (remaining demand, overtime production capacity)].

*Step 5.3.3.* The number of deliveries and delivered quantity in each period is determined (the details will be explained in the subsequent part of the paper).

*Step 5.4.* The generated nodes add to the *open-list*, and the parent (start node) is added to the *closed-list* and the parent's costs (f, g, and h) are memorized.

*Step 5.5.* If the number of members in the *open-list* exceeds the default number, the default number of nodes with the lowest cost is memorized in the *open-list* and others are ignored.

*Step 5.6.* The *f(n)* cost for each node in the open-list is calculated. The node with the lowest cost is selected, and the parent node is replaced with this node. Branching is done on this new parent.

*Step 5.7.* If a node with zero demand (total delivery is equal to the total demand) is added to the *closed-list* (as the lowest cost node in the open-list), then the goal state is achieved and the algorithm is terminated; otherwise, repeat steps 5.3 to 5.7.

## 3.3. Evaluation Functions for the A* Search

The $g(n)$ and $h(n)$ cost functions are two evaluation functions in A* search which are calculated according to the objective function in the lower level decision-making model. The $h(n)$ function never overestimates the cost of achieving the goal state (optimistic cost function). For example, the estimation of production cost for next periods optimistically calculates through considering the ordinary cost (instead of overtime cost) for overtime productions as a lower bound. In fact, the heuristic function estimates the future costs less than it is, so this heuristic is *admissible*. The solution representation for the lower-level decision-makers is shown in Fig. 3:

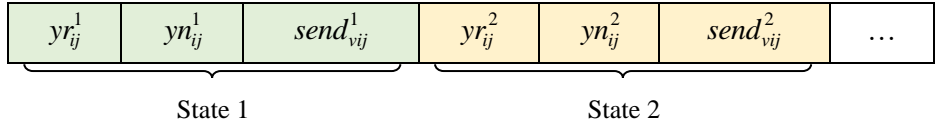

**Figure 3.** The representation of an A* solution.

The calculations associated with the proposed algorithm are presented in the Appendix. Some essential points in the calculation of the heuristic cost function are as follows: (1) The inventory cost is not considered between two consecutive periods, (2) the ordinary production cost is considered for overtime production (instead of overtime production cost). The variable values, which are supposed to be sent to the upper-level model, are determined according to the proposed procedure by finding the desired state. A flowchart of the A* search algorithm for finding the best proposals for each supplier is provided in Fig. 4. Also, the flowchart of the negotiation is based on bi-level programming and the PSO algorithm in Fig. 5 (Kaheh *et al.*, 2018).

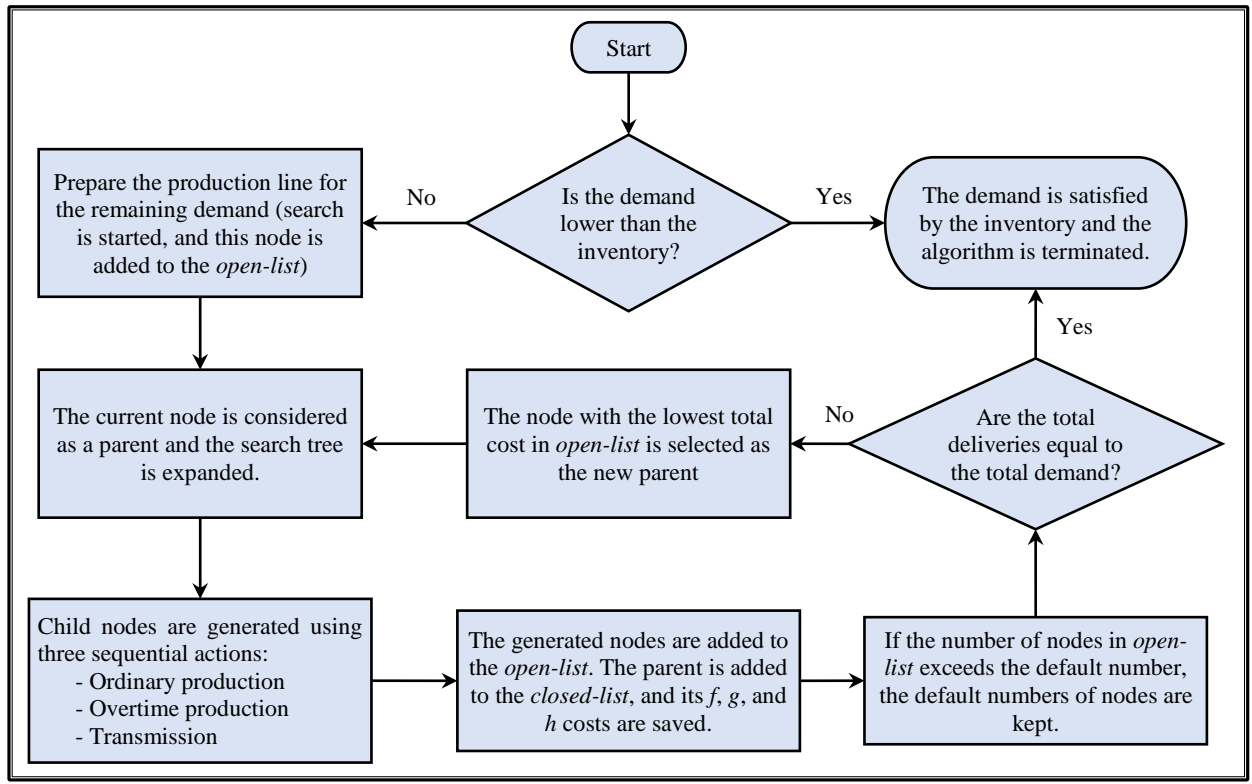

**Figure 4.** Flowchart of the A* search algorithm for finding the best proposals for each supplier.

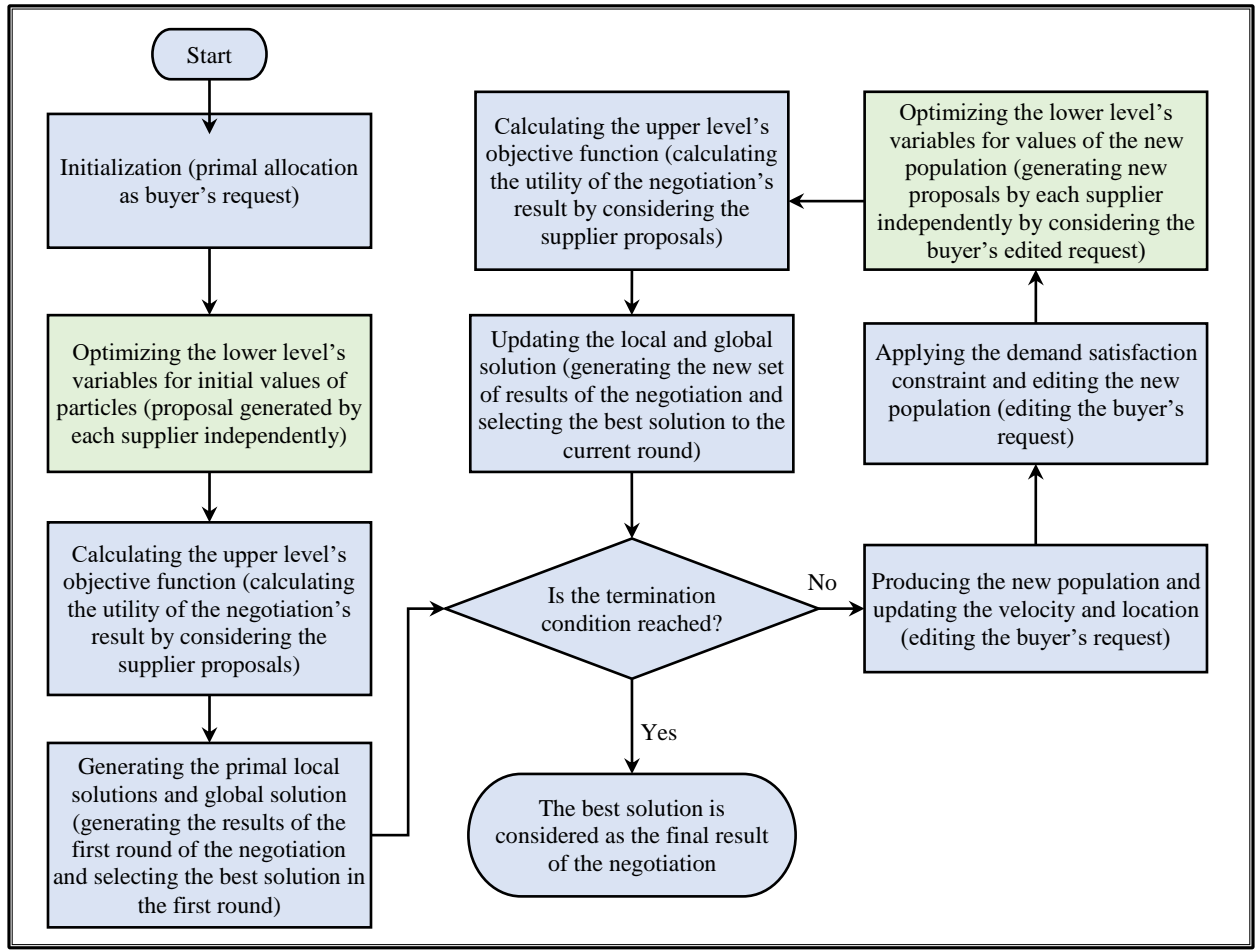

**Figure 5.** Flowchart of the negotiation based on bi-level programming and PSO algorithm.

## 4. A Case Study

Our research is established based on a supplying automotive parts company (SAPCO), which provides the required parts for one of the largest automotive manufacturers in Iran. Thus, SAPCO has an essential role to manage transactions with multiple suppliers, and maintain valuable partnerships in the automotive supply chain. In general, this research has been conducted based on the assumptions derived from the interviews with the experts in SAPCO and their partners. Moreover, our proposed algorithm was carried out with the real data collected in these companies. The collected data are related to 25 glass types for different cars that are supplied by 5 local suppliers.

### 4.1. Computational Analysis

This section contains the computational results of the proposed algorithm that was practically applied to solve our bi-level programming problem. All algorithms have been implemented as a computer program in MATLAB software and executed on a Core i5 3.2 GHz processor with 4 GB of main memory. To illustrate the main outputs of the PSO-A* algorithm, the results of two hypothetical examples solved through PSO-A* are summarized in Tables 3 and 4. To evaluate the performance of the PSO-A* algorithm, the computational results of the PSO-A* mechanism for several problems are compared with two other algorithms called PSO-Greedy and PSO-Simulated Annealing (PSO-SA). The greedy search finds the desired state based on the estimated cost through the proposed heuristic function ($h(n)$) and prefers to follow a single path, without considering whether this will be best in the long run. The heuristic function used for greedy search is similar to that of the A* search.

**Table 3.** The results of the PSO-A* algorithm for a hypothetical example with one item and 5 suppliers with $q_j = 1150$.

| Suppliers | $Q_i^{min}$ | $Q_i^{max}$ | $q_i$ | $p_i$ | Upper-level objective value | |
|---|---|---|---|---|---|---|
| | | | | | $w_1 = 0.4, w_2 = 0.6$ | $w_1 = 1, w_2 = 0$ |
| 1 | 180 | 500 | 201 | 6622.862 | | |
| 2 | 150 | 500 | 221 | 3292.537 | | |
| 3 | 150 | 500 | 150 | 10932.740 | 1071478 | 2676542 |
| 4 | 200 | 500 | 426 | 4331.018 | | |
| 5 | 150 | 500 | 152 | 9501.610 | | |

The typical drawback of heuristic algorithms is their inability to escape local optima. However, it is possible to remedy that by embedding various mechanisms to escape local optima. To evaluate the proposed algorithm, the output of the PSO-A* algorithm with those of the hybrid PSO-Simulated annealing algorithm has been compared. Simulated annealing (SA) is a local search algorithm, which avoids the local optimum by accepting a non-improving neighboring solution with a probability; the initial solution of SA is generated from greedy search, also the neighboring solution is generated based on a random selection from the feasible space. The initial temperature is set large enough such that almost all the transitions are accepted in the initial stages.

**Table 4.** The results of the PSO-A* algorithm for a hypothetical example with 5 items and 3 suppliers with $q = 500, 500, 500, 1000, 800$.

| Item | $Q_{1j}^{min}$ | $Q_{2j}^{min}$ | $Q_{3j}^{min}$ | $Q_{1j}^{max}$ | $Q_{2j}^{max}$ | $Q_{3j}^{max}$ | $q_{1j}$ | $q_{2j}$ | $q_{3j}$ | $p_{1j}$ | $p_{2j}$ | $p_{3j}$ | Upper-level objective value | |
|---|---|---|---|---|---|---|---|---|---|---|---|---|---|---|
| | | | | | | | | | | | | | $w_1 = 0.4, w_2 = 0.6$ | $w_1 = 1, w_2 = 0$ |
| 1 | 100 | 110 | 110 | 250 | 300 | 300 | 193 | 172 | 135 | 2885.2 | 2947.6 | 3814.8 | | |
| 2 | 120 | 130 | 130 | 250 | 300 | 300 | 187 | 130 | 183 | 1777.6 | 2537.3 | 1793.5 | | |
| 3 | 140 | 150 | 150 | 250 | 300 | 300 | 140 | 221 | 279 | 3057.3 | 2266.9 | 1931.9 | 3339497 | 8348068 |
| 4 | 145 | 155 | 155 | 500 | 500 | 500 | 241 | 277 | 482 | 2075.6 | 1719.5 | 1039.8 | | |
| 5 | 145 | 155 | 155 | 300 | 500 | 500 | 156 | 387 | 257 | 19788 | 5640.8 | 8630.0 | | |

Each of PSO-A*, PSO-Greedy, and PSO-SA algorithms is carried out ten times to solve each sample problem. The deviation of solutions from the best-found solution (for each problem through three algorithms) is calculated according to (33). It should be stated that the parameters of these algorithms are set up in an appropriate setting by performing several experiments.

$$dev_{PSO\_A^*} = \frac{sol_{PSO\_A^*} - sol_{best\_found}}{sol_{best\_found}}; \qquad dev_{PSO\_Exact} = \frac{sol_{PSO\_exact} - sol_{best\_found}}{sol_{best\_found}}$$

$$dev_{PSO\_Greedy} = \frac{sol_{PSO\_greedy} - sol_{best\_found}}{sol_{best\_found}}; \qquad dev_{PSO\_SA} = \frac{sol_{PSO\_SA} - sol_{best\_found}}{sol_{best\_found}} \qquad (33)$$

$$sol_{best\_found_1} = \underset{k}{Min}\{sol_{PSO\_A^*}, sol_{PSO\_Exact}, sol_{PSO\_Greedy} \mid k : \text{test\_number}\}$$

$$sol_{best\_found_2} = \underset{k}{Min}\{sol_{PSO\_A^*}, sol_{PSO\_SA}, sol_{PSO\_Greedy} \mid k : \text{test\_number}\}$$

Since different parameters and factors affect the performance of a hybrid algorithm, choosing the best combination of the parameters can intensify the search process and prevent premature convergence (Hamta, *et al.* 2013). Therefore, based on several experiments, in which the algorithm is carried out with different levels of algorithm parameters (Table 5), in which the best value of each parameter is indicated in bold.

**Table 5.** Factors and their corresponding levels for the PSO-A* algorithm.

| Number of iterations | | | | Number of particles | | Cognitive coefficient ($c_1$) | | | Social coefficient ($c_2$) | | | $W_{min}$ | | | $W_{max}$ | | |
|---|---|---|---|---|---|---|---|---|---|---|---|---|---|---|---|---|---|
| 1 | 2 | 3 | 4 | 1 | 2 | 1 | 2 | 3 | 1 | 2 | 3 | 1 | 2 | 3 | 1 | 2 | 3 |
| 80 | 90 | **100** | 110 | 20 | **30** | 1.5 | **2** | 2.5 | 1.5 | 2 | **2.5** | **0.1** | 0.2 | 0.3 | 0.7 | **0.8** | 0.9 |

The averages of deviations for each problem in all implementations through each algorithm are brought in Table 6. It can be concluded that PSO-A* algorithm is more effective compared with the two other algorithms. Although the runtime of the PSO-Greedy is mostly less than that of PSO-A*, the deviations of PSO-A* solutions are significantly less than the PSO-Greedy, and consequently, PSO-A* is more effective than PSO-Greedy. The comparisons among three algorithms for large-size problems are shown in Table 7.

**Table 6.** Comparison between the deviations from the best-found solution for small-size problems.

| Problem No. | # of suppliers | # of items | Exact | | PSO-Greedy | | PSO-A* | |
|---|---|---|---|---|---|---|---|---|
| | | | Average Deviation (%) | Average runtime (s) | Average Deviation (%) | Average runtime (s) | Average Deviation (%) | Average runtime (s) |
| 1 | 2 | 1 | 0 | 739.356 | 0.09 | 39.761 | 0.04 | 106.222 |
| 2 | 2 | 2 | 0 | 941.219 | 0.23 | 56.291 | 0.06 | 175.811 |
| 3 | 2 | 3 | 0 | 1341.510 | 0.19 | 71.732 | 0.00 | 251.930 |
| 4 | 2 | 5 | 0 | 1789.864 | 0.35 | 106.541 | 0.01 | 418.643 |
| 5 | 2 | 7 | 0 | 2988.494 | 0.21 | 138.511 | 0.00 | 579.781 |
| 6 | 3 | 1 | 0 | 1089.965 | 0.20 | 51.192 | 0.04 | 218.118 |
| 7 | 3 | 2 | 0 | 1780.312 | 0.28 | 73.529 | 0.00 | 296.397 |
| 8 | 3 | 3 | 0 | 6321.341 | 0.25 | 98.543 | 0.08 | 379.331 |
| 9 | 3 | 5 | 0 | 13741.782 | 0.43 | 131.984 | 0.00 | 550.528 |
| 10 | 3 | 7 | 0 | 38954.910 | 0.54 | 169.631 | 0.00 | 718.459 |
| 11 | 4 | 1 | 0 | 1341.631 | 0.27 | 71.620 | 0.06 | 349.762 |
| 12 | 4 | 2 | 0 | 2385.612 | 0.51 | 99.731 | 0.00 | 441.451 |
| 13 | 4 | 3 | 0 | 9941.139 | 0.49 | 116.411 | 0.02 | 596.561 |
| 14 | 5 | 1 | 0 | 1834.561 | 0.31 | 89.819 | 0.00 | 437.797 |
| Average | | | 0 | – | 0.31 | – | 0.02 | – |

The PSO-A*, PSO-Greedy, and PSO-SA for large-size problems have been conducted with random data. Random data have been generated by fitting the appropriate distribution function to the collected data. For example, to produce the random data for the demand parameter, a uniform distribution in the range of 300 to 1000 has been applied; also the production time is generated from a uniform distribution in the range of 3 to 5.5, and so on. As shown in Table 7, the solution of the PSO-A* algorithm is superior to those of the PSO-SA algorithm. In general, the proposed PSO-A* algorithm outperformed other algorithms.

Table 7. Comparison between the results of PSO-A*, PSO-SA, and PSO-Greedy for large-size problems.

| Problem No. | # of suppliers | # of items | PSO-SA Average Deviation (%) | PSO-Greedy Average Deviation (%) | PSO-A* Average Deviation (%) |
|---|---|---|---|---|---|
| 1 | 8 | 30 | 0.03 | 0.35 | 0.00 |
| 2 | 8 | 50 | 0.00 | 0.47 | 0.03 |
| 3 | 10 | 50 | 0.18 | 0.34 | 0.00 |
| 4 | 10 | 70 | 0.00 | 0.37 | 0.01 |
| 5 | 15 | 70 | 0.13 | 0.39 | 0.00 |
| 6 | 15 | 80 | 0.03 | 0.45 | 0.00 |
| 7 | 20 | 80 | 0.19 | 0.28 | 0.00 |
| 8 | 20 | 100 | 0.08 | 0.43 | 0.00 |
| Average | - | - | 0.08 | 0.38 | 0.005 |

In addition, we performed a sensitivity analysis for different weights of objective functions and different $\gamma$ (delay adjustment coefficient), the results of which are reported in Fig. 6. The sensitivity analysis demonstrates that: (1) for small amounts of $w_1$, especially for $w_1$= 0, 0.1, and 0.2 and the values of gamma between 0.8 and 0.9 the buyer's cost decreases as the reduction of delay cost exceeds the increment of the procurement cost. It is because by increasing the value of $\gamma$, the suppliers' delay cost increases, so they prefer to use their overtime capacity, and thereby their operational cost and proposed prices will magnify. On the other hand, the buyer's shortage cost, which is affected by the suppliers' delay, is more important than procurement cost, so the concentration of suppliers for avoiding the delay raises the buyer's cost.

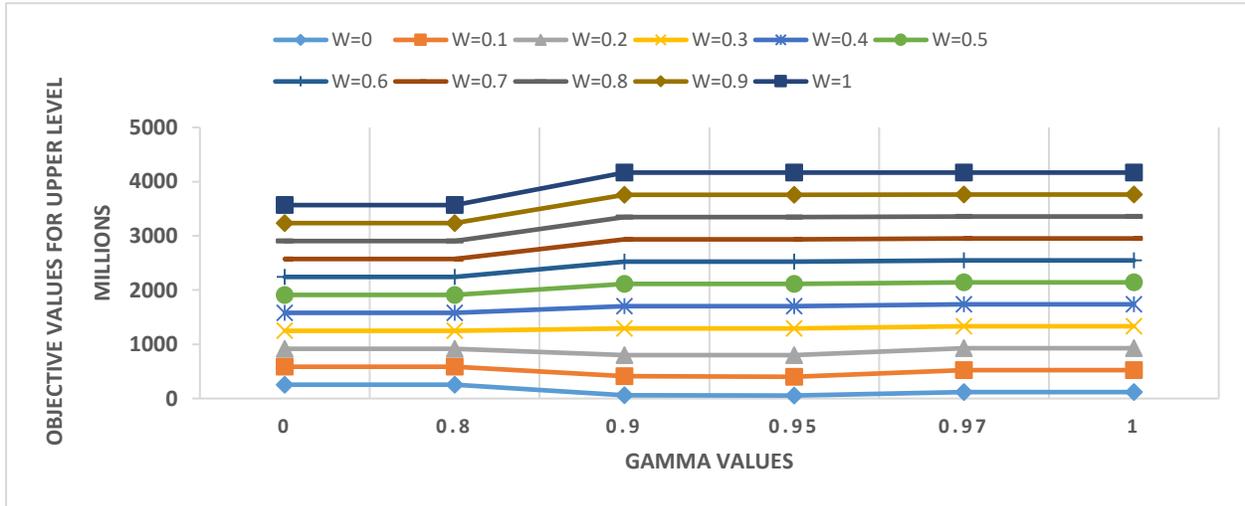

Figure 6. Sensitivity analysis of the model for different weight values of each objective and gamma coefficient.

After a stable state, suggesting the balance between the delay cost and procurement cost, there is a slight growth in buyer cost. This is because the development of procurement cost exceeds the reduction in delay cost for the gamma values between 0.95 and 0.97. It can be concluded that the suppliers' attention to avoid delays is beyond the importance of delay for the buyer. (2) regarding the large value of $w_1$, especially the values of first weight between $w_1$ = 0.5 to 1, by increasing the $\gamma$ values between 0.8 and 0.9 the buyer's cost increases and the reduction of delay cost will not be significant. This is attributed to the fact that for $w_1$ =

0.5 to 1 the delay cost loses its importance for the buyer so the suppliers' attention to delay avoidance magnifies the buyer's cost. (3) For $w_1 = 0.3$ and 0.4, the buyer's cost is not sensitive to the gamma values.

The percentage of average price reduction, influenced by the reduction of suppliers' operational costs is provided in Table 8. As shown in this table, the prices, as one of the PSO-A* algorithm outputs, are compared with the collected price data for the number of items supplied by five suppliers. The results demonstrate that applying our proposed algorithm for solving the real-world procurement problem is satisfactory, and it could be applied for similar procurement problems.

**Table 8.** The percentage of average cost reduction.

| Item category | Avg. price reduction | Item category | Avg. price reduction | Item category | Avg. price reduction |
|---|---|---|---|---|---|
| 1 | 39% | 4 | 41% | 7 | 29% |
| 2 | 44% | 5 | 38% | 8 | 43% |
| 3 | 32% | 6 | 36% | 9 | 31% |

## 5. Conclusions and Future Research

This paper addresses a distributed decision-making process in a procurement problem designed based on bi-level programming. In decentralized systems, generally, achieving a near-optimal solution that promotes the parties to gain the agreement is more preferable to the optimal solutions for each partner separately. Also, by taking into account the suppliers' production planning in order allocation, the inventory and delay costs will reduce and last-minute orders will be avoided, which are beneficial to all parties. In this approach, negotiated parties, based on the nature of the bi-level programming can reform their decision variable values while considering the other parties' constraints, without any strict controls.

We presented a bi-level nonlinear mixed-integer programming for a real-world problem in an automotive part supplier company. Typical negotiation components perfectly match the bi-level programming model and its solution procedure. The results have shown that the PSO-A* algorithm is more effective compared to PSO-Greedy and PSO-SA algorithms. The output of this study can be applied as a pre-negotiation tool in a real-world procurement problem, and the model provides the parties with the opportunity to achieve near-optimal solutions that enhance their bargaining possibilities.

The advantages of our presented method are as follows:

1. Despite the development of mathematical models to capture the negotiation process in many papers (e.g. Cheng, 2011 & Jung *et al.*, 2008), there has been no strict definition of negotiation structures in mathematical models. In this paper, typical negotiation components perfectly match the mathematical model and its solution procedure. Compared to other bi0level models, the current work considers the most comprehensive cost types as illustrated in Table 1.

2. The developed negotiation mechanism presents a win-win game, which makes the partners follow their objectives. In other words, the model provides the parties with the opportunity to achieve near-optimal solutions, which enhances their bargaining possibilities. This feature satisfies the partners and supports the partnership with valued suppliers.

3. An innovative hybrid algorithm for hierarchical distributed procurement problems is presented, which performed quite well for solving a real procurement case.

The practical applications of this research are as follows:

1. The buyer (auctioneer) can use the output of this paper as a pre-negotiation tool. The auctioneer can prepare appropriate requests about the quantity and due date through this simulated negotiation. Also, through this mechanism, the interests of each supplier are satisfied.
2. Because of the non-cooperative nature of the decision-making among the partners with different levels of bargaining power, the partnership may be unstable. This issue increases the cost in the supply chain and damage all partners. Especially, because of some mismanagement in the Iranian automotive supply chain or similar countries, costs are too high and the consequences of this occurrence are high prices and pressure on the consumers. Therefore, by applying this mechanism, the costs have been reduced by integrating the partners' conflicting interests.
3. Since time is a valuable resource for managers in dynamic systems, a negotiation mechanism with appropriate time is an essential requirement. The proposed mechanism makes partners reach an agreement in a reasonable time.

Possible extensions of this research are the following: (1) improving the mathematical model by considering more details such as suppliers' scheduling, transportation planning, and advanced pricing mechanisms. (2) Embedding some learning mechanisms to use the previous elite proposals for next periods in a dynamic multi-period contract. (3) Considering the exceptions and abnormal situations in procurement such as supplier negligence to produce certain items in some periods. (4) Comparing the proposed algorithm with other metaheuristic algorithms.

# References

Akartunali, K., & Miller, A. J. (2009). A heuristic approach for big bucket multi-level production planning problems. *European Journal of Operational Research* 193(2): 396-411.

Baradaran Kazemzadeh, R., Kaheh, Z., & Masehian, E. (2014). A Mixed Integer Nonlinear Programming Model for Order Replenishment and a Heuristic Algorithm for its Solution. *Journal of Industrial Engineering Research in Production Systems*, *2*(3), 63-75.

Ben-Ayed, O., & Blair, C.E. (1990). Computational difficulties of bi level linear programming. *Operations Research* 38, 556–560.

Carrión, M., Arroyo, J.M., & Conejo, A.J. (2009). A bi level stochastic programming approach for retailer futures market trading. *IEEE Transactions on Power Systems* 24, 1446–1456.

Chalmardi, M. K., & Camacho-Vallejo, J. F. (2019). A bi-level programming model for sustainable supply chain network design that considers incentives for using cleaner technologies. Journal of Cleaner Production, 213, 1035-1050.

Cheng, C.-B. (2011). Reverse auction with buyer-supplier negotiation using bi-level distributed programming. *European Journal of Operational Research* 211(3): 601-611.

Cheng, G., Zhao, S., & Zhang, T. (2019). A Bi-Level Programming Model for Optimal Bus Stop Spacing of a Bus Rapid Transit System. Mathematics, 7(7), 625.

Danna, E., Rothberg, E., & Le Pape, C. (2005). Exploring relaxation induced neighborhoods to improve MIP solutions. Mathematical Programming 102, 71–90.

Fang, F. & Wong, T.N. (2010). Applying hybrid case-based reasoning in agent-based negotiations for supply chain management. *Expert Systems with Applications* 37(12): 8322-8332.

Hart, P. E., Nilsson, N. J., & Raphael, B. (1968). A formal basis for the heuristic determination of minimum cost paths. *Systems Science and Cybernetics, IEEE Transactions on*, *4*(2), 100-107.

Hamta, N., Fatemi Ghomi, S. M. T., Jolai, F., & Akbarpour Shirazi, M. (2013). A hybrid PSO algorithm for a multi-objective assembly line balancing problem with flexible operation times, sequence-dependent setup times and learning effect. *International Journal of Production Economics*, *141*(1), 99-111.

Hansen, E. A., & Zhou, R. (2007). Anytime heuristic search. Journal of Artificial Intelligence Research, 28, 267–297.

Hejazi, S. R., Memariani, A., Jahanshahloo, G., & Sepehri, M. M. (2002). Linear bilevel programming solution by genetic algorithm. *Computers & Operations Research*, *29*(13), 1913-1925.

Jang, J., & Do Chung, B. (2020). Aggregate production planning considering implementation error: A robust optimization approach using bi-level particle swarm optimization. Computers & Industrial Engineering, 142, 106367.

Jia, Z. Z., Deschamps, J. C., & Dupas, R. (2016). A negotiation protocol to improve planning coordination in transport-driven supply chains. *Journal of Manufacturing Systems*, *38*, 13-26.

Jiang, Y., Li, X., Huang, C., & Wu, X. (2013). Application of particle swarm optimization based on CHKS smoothing function for solving nonlinear bilevel programming problem. *Applied Mathematics and Computation*, *219*(9), 4332-4339.

Jung, H., Jeong, B., & Lee, C. G. (2008). An order quantity negotiation model for distributor-driven supply chains. *International Journal of Production Economics*, *111*(1), 147-158.

Kadadevaramath, R. S., Chen, J. C., Latha Shankar, B., & Rameshkumar, K. (2012). Application of particle swarm intelligence algorithms in supply chain network architecture optimization. *Expert Systems with Applications*, *39*(11), 10160-10176.

Kaheh, Z., Kazemzadeh, R. B., Masehian, E., Hosseinzadeh Kashan, A. (2018) Developing A Bi-Level Programming Model for Procurement Management and A Hybrid Algorithm for Its Solution, Volume 33.1, Issue 2.1, 21-33.

Kaheh, Z., Kazemzadeh, R. B., & Sheikh-El-Eslami, M. K. (2019). Simultaneous consideration of the balancing market and day-ahead market in Stackelberg game for flexiramp procurement problem in the presence of the wind farms and a DR aggregator. IET Generation, Transmission & Distribution, 13(18), 4099-4113. (1)

Kaheh, Z., Baradaran Kazemzadeh, R., & Sheikh-El-Eslami, M. K. (2019). A trilevel programming model for flexiramp and reserve procurement in high penetration of wind farms and participation of a large industry and a DR aggregator. International Transactions on Electrical Energy Systems, 29(7), e12105. (2)

Kaheh, Z., Baradaran Kazemzadeh, R., & Sheikh-El-Eslami, M. K. (2020). A solution based on fuzzy max-min approach to the bi-level programming model of energy and exiramp procurement in day-ahead market. *Scientia Iranica*, *27*(2), 846-861.

Baradaran Kazemzadeh, R., Kaheh, Z., & Masehian, E. (2014). A Mixed Integer Nonlinear Programming Model for Order Replenishment and a Heuristic Algorithm for its Solution. *Journal of Industrial Engineering Research in Production Systems*, *2*(3), 63-75.

Kuo, R. J. & Han, Y. S. (2011). A hybrid of genetic algorithm and particle swarm optimization for solving bi-level linear programming problem - A case study on supply chain model. *Applied Mathematical Modelling* 35(8): 3905-3917.

Ma, W., Wang, M., & Zhu, X. (2013). Hybrid particle swarm optimization and differential evolution algorithm for bi-level programming problem and its application to pricing and lot-sizing decisions. Journal of Intelligent Manufacturing, 1-13.

Moiseeva, E., & Hesamzadeh, M. R. (2017). Strategic bidding of a hydropower producer under uncertainty: Modified benders approach. IEEE Transactions on Power Systems, 33(1), 861-873.

Proch, M., Worthmann, K., & Schlüchtermann, J. (2017). A negotiation-based algorithm to coordinate supplier development in decentralized supply chains. *European Journal of Operational Research*, *256*(2), 412-429.

Ocampo, L. A., Vasnani, N. N., Chua, F. L. S., Pacio, L. B. M., & Galli, B. J. (2021). A bi-level optimization for a make-to-order manufacturing supply chain planning: a case in the steel industry. *Journal of Management Analytics*, 1-24.

Qu, B. Y., Liang, J. J., & Suganthan, P. N. (2012). Niching particle swarm optimization with local search for multi-modal optimization. *Information Sciences*, *197*, 131-143.

Roghanian, E., Aryanezhad, M. B., & Sadjadi, S. J. (2008). Integrating goal programming, Kuhn–Tucker conditions, and penalty function approaches to solve linear bi-level programming problems. *Applied Mathematics and Computation*, *195*(2), 585-590.

Russell,S & Norvig,P. *Artficial Intelligence: A Modern Approach*. Prentice-Hall, Saddle River, NJ, (1995).

Sakawa, M. & Matsui, T. (2013). Interactive random fuzzy two-level programming through possibility-based probability model. *Information Sciences* 239(0): 191-200.

Sandholm, T., Suri, S., Gilpin, A., & Levine, D., (2002). Winner determination in combinatorial auction generalizations. In *Proceedings of the first international joint conference on Autonomous agents and multiagent systems: part 1* (pp. 69-76). ACM.

Sedighizadeh, D., & Masehian, E. (2009). Particle swarm optimization methods, taxonomy and applications. *international journal of computer theory and engineering*, *1*(5), 486-502.

Soares, I., Alves, M. J., & Antunes, C. H. (2019, July). A bi-level hybrid PSO: MIP solver approach to define dynamic tariffs and estimate bounds for an electricity retailer profit. In Proceedings of the Genetic and Evolutionary Computation Conference Companion (pp. 33-34).

Shokr, I., & Torabi, S. A. (2017). An enhanced reverse auction framework for relief procurement management. *International journal of disaster risk reduction*, *24*, 66-80.

Tate, W. L., Ellram, L. M., & Dooley, K. J. (2012). Environmental purchasing and supplier management (EPSM): Theory and practice. *Journal of Purchasing and Supply Management*, *18*(3), 173-188.

Wan, Z., Wang, G., & Sun, B. (2013). A hybrid intelligent algorithm by combining particle swarm optimization with chaos searching technique for solving nonlinear bilevel programming problems. *Swarm and Evolutionary Computation*, *8*, 26-32.

Wu, C. and Barnes, D. (2011). A literature review of decision-making models and approaches for partner selection in agile supply chains. *Journal of Purchasing and Supply Management* 17(4): 256-274.

Zhang, R., & Wang, K. (2019). A multi-Echelon global supply chain network design based on transfer-pricing strategy. *Journal of Industrial Integration and Management*, *4*(01), 1850020.24

# Appendix

To compute the quantity in each delivery, the inventory cost and the delivery cost are considered to be equal (34). It should be noted that $t'$ (the time interval between two sequential periods) is not considered in this case.

$$\frac{H_{ij}PT_{ij}(send_{vij}^t)^2}{2} = \alpha + \beta(send_{vij}^t) \rightarrow \frac{H_{ij}PT_{ij}}{2}(send_{vij}^t)^2 - \beta(send_{vij}^t) - \alpha = 0 \qquad (34)$$

So the allowed quantity in each delivery without interval time is calculated through

$$allowed\_quantity(aq_{ij}) = \min\left\{inventory\_capacity, vehicle\_capacity, remaining\_demand, \frac{\beta + \sqrt{\beta^2 + 2\alpha H_{ij}PT_{ij}}}{H_{ij}PT_{ij}}\right\}. \qquad (35)$$

In this way, the number of deliveries is calculated using:

$$\min\left\{\text{available vehicles in each period}, \frac{total\ production\ in\ each\ period}{allowed\ quantity\ in\ per\ transmission}\right\} \qquad (36)$$

The previous case provides a lower bound for each delivery volume and the number of vehicles. Now, the case in which the interval time between two consecutive periods (and inventory cost during this time) is taken into account; and thereby, the inventory cost will be higher at the end of the period. Therefore, delivering the leftover inventory is more preferable to holding it, and as a result, the delivered quantity in each period increases as well. Since the unused inventory at the end of each period could be calculated through mentioned relations, it is possible to equally distribute it among in-use vehicles during the same period. In this circumstance, there are three sequential decisions: (1) distributing the inventory among in-use vehicles. (2) Using extra vehicles, in the case that the current vehicles are not enough. (3) Holding the remaining inventory as long as the two previous decisions are not enough for delivering the leftover inventory.

To calculate the optimized augmented value to each delivery, the inventory cost and delivery cost will be equalized (37). The augmented value may not be the coefficient of the number of vehicles, so its quotient ($x_{ij}$) adds to each delivery and its remaining redistribute among vehicles. (Fig. 4)

$$\left\{nv_{use}\underbrace{\left[\frac{(aq_{ij}+x_{ij})^2 \times PT_{ij} \times H_{ij}}{2}\right]}_{1} + \underbrace{\left[\frac{(I_{ij}^t - (nv_{use} \times x_{ij}))^2 \times PT_{ij} \times H_{ij}}{2}\right]}_{2} + \underbrace{[(I_{ij}^t - (nv_{use} \times x_{ij})) \times H'_{ij}]}_{3} - \underbrace{[\frac{(I_{ij}^{t-1})^2 \times PT_{ij} \times H_{ij}}{2}]}_{4}\right\} \qquad (37)$$
$$= \alpha(nv_{use}) + \beta(nv_{use} \times (aq_{ij} + x_{ij}))$$

The total allowed volume delivered in each period is calculated using (38) by considering the time interval between two sequential periods:

$$Total\ allowed\ quantity\ for\ delivery = \min\{(inventory \times number\ of\ used\ vehicles), (vehicle\ capacity \times used\ vehicles) + [yr_{ij}^t + yn_{ij}^t + I_{ij}^{t-1}], [number\ of\ used\ vehicles \times (aq_{ij} + x_{ij})]\} \qquad (38)$$

The remaining inventory is either delivered by additional vehicles or stocked. It is important to note that the time interval between two sequential periods and initial inventory in each period are not considered in the heuristic cost function ($h(n)$), but it is considered in cost from the start node to the current node ($g(n)$). In total, for calculating the $g(n)$ cost, the inventory cost, delivery cost, production cost, setup cost, and delay cost should be considered according to the lower-level objective function.

The following factors must be considered to calculate the $h(n)$ cost:

(1) *TFS*: The total quantities delivered in future periods are calculated according to the remaining demand and the previous period inventory.

(2) *TFV*: The number of future deliveries which is calculated through Eq. 39.

$$TFV = \min \left\{ \frac{\text{total quantity to be delivered}}{\text{vehicles capacity}} ; \frac{\text{total quantity to be delivered}}{\text{inventory capacity}} \right\} \tag{39}$$

(3) *TFD*: The number of future days to fulfill the orders according to Eq. 40.

$$TFD = \max \left\{ \frac{\text{number of future delivery}}{\text{available vehicles in each period}} ; \frac{\text{processing time of remaining production}}{\text{ordinary time and overtime capacity}} \right\} \tag{40}$$

(4) *FDS*: The daily delivery, and (5) FD: The number of future working days. Thus, the h cost is calculated through Eq. 41:

$$h(n) = (cor_{ij} \cdot FP_{ij}) + (FD \cdot sc_{ij}) + \alpha[TFV] + \beta(TFS) + \frac{[TFV] \cdot h_{ij} \cdot H_{ij} \cdot FAS}{2} + (delay\_cost) \tag{41}$$

If the current period plus future working day runs out of the announced due date. The delay cost is calculated through Eq. 42.

$$\begin{aligned} delay\_cost &= \gamma \cdot \sum_{i=current-period}^{last-period} FDS \cdot i = \gamma \cdot FDS \sum_{i=current-period}^{last-period} i \\ current\_period &= \max\{1, t+1 - LT_{lower}\} \\ last\_period &= [t + TFD - LT_{lower}] \end{aligned} \tag{42}$$